\documentclass[mathleft,fleqn,%
]{an}
\usepackage{natbib}
\usepackage{graphicx}
\usepackage[varg]{txfonts}
\begin{document}

\Pagespan{1}{}
\Yearpublication{2014}%
\Yearsubmission{2014}%
\Month{0}%
\Volume{999}%
\Issue{0}%
\DOI{asna.201400000}%

\title{X-ray variability of cool stars: 
        Magnetic activity and accretion} 

\author{B. Stelzer\inst{1}\fnmsep\thanks{Corresponding author:
        {stelzer@astropa.inaf.it}}}
\titlerunning{X-ray variability of cool stars}
\authorrunning{B. Stelzer}
\institute{
INAF - Osservatorio Astronomico di Palermo, Piazza del Parlamento 1, 90134 Palermo, Italy}

\received{XXXX}
\accepted{XXXX}
\publonline{XXXX}


\abstract{%
  This article provides a review of X-ray variability from late-type stars with 
  particular focus on the achievements of {\em XMM-Newton} and its potential for future 
  studies in this field. 
}
\maketitle

\section{X-ray emission from cool stars}\label{sect:intro} 

One of the most prominent signatures of the X-ray emission observed from late-type stars is
its variability. In analogy to the Sun, the bulk of the X-ray emission from cool 
stars is a manifestation of magnetic activity, i.e. a result of the stellar dynamo which 
continuously regenerates the magnetic field against dissipation. It is, thus, a genuinely 
dynamic process.  

Since stellar dynamos require the presence of 
a convection zone, they are believed to operate only in the cooler half of the
Hertzsprung-Russell diagram, in stars of spectral types F and later. 
The second key ingredient of stellar dynamos is (differential) rotation. 
As a consequence, in the faster rotating, younger stars magnetic activity is scaled up 
with respect to main-sequence (MS) 
stars, e.g. $10^3$ times higher X-ray luminosity is detected from pre-main sequence 
(pre-MS) {\it T\,Tauri stars (TTS)} 
with respect to MS dwarfs \citep{Stelzer01.1, Preibisch05.1}. 

Similar to the Sun and late-type MS stars in general, 
TTS display various kinds of magnetic activity phenomena. 
In addition, their X-ray emission and its variability may carry the signatures of 
various kinds of interaction with their circumstellar environment. 
In fact, during the first few Myrs of their evolution 
late-type stars pass through several phases that are
characterized by the amount of circumstellar material and its 
interaction with the forming star. 
While there are no certain reports on X-ray emission from the earliest protostellar stages 
when the star is surrounded by an optically thick envelope and radiates mostly 
at sub-mm and longer wavelengths (so-called {\em Class\,0 object}),
X-ray emission was seen from several of the somewhat more evolved {\em Class\,I protostars}.
However, the X-ray properties (luminosity, plasma temperature and column density) of 
protostars have remained poorly constrained so far \citep{Prisinzano08.0, Pillitteri10.1}. 
In the subsequent evolutionary phase ({\em Class\,II objects}) the 
envelope has dispersed and settled into a disk from which the star
generally accretes matter ({\em classical TTS}). Thanks to the superb spectral 
capabilities of {\em XMM-Newton} and {\em Chandra} a soft high-density X-ray component 
was identified in a small number of bright, accreting T\,Tauri stars which was ascribed
to shocks at the base of the accretion flow \citep{Kastner02.1, Stelzer04.3}.
Moreover, the spatial and spectral properties of the X-ray emission observed from 
some young stars have revealed X-ray emission from shocks in protostellar jets 
\citep{Guedel07.1}. 
When finally all circumstellar matter is gone and only the 
{\em Class\,III object}\footnote{See e.g. \cite{Feigelson99.1} for details on the 
classification scheme of pre-main sequence stars.} 
or {\em weak-line TTS} is left, the viable X-ray emission mechanisms reduce to 
those based on (scaled-up) magnetic activity. 

In the remainder of this article I review
recent developments in studies of X-ray variability from late-type stars focusing
on magnetic activity and accretion.

\section{Activity-related X-ray variability}\label{sect:activity}

Stellar magnetic activity can produce variable X-ray emission on a vast range of time-scales
from minutes to decades. Drastic brightness increases and subsequent decay to the 
previous emission level with typical time-scales of hours 
are the most prominent activity signature. These 
{\em flares} represent a rapid release of magnetic energy through reconnection events.  
At the long end of the variability time-scales are the {\em dynamo cycles}, analogs of the 
solar $11$-year cycle which is characterized
by a periodic variation of the overall X-ray output of the Sun by a factor $\sim 20...100$
depending on the energy band \citep{Orlando01.1}.
While flares and cycles ultimately go back to intrinsic variability of the magnetic field,
there are also variations related to geometric effects. In particular, inhomogeneous
surface coverage with magnetically active regions together with surface rotation 
can be expected to lead to periodically modulated X-ray emission reflecting the
time-scale of stellar rotation. 
In the following, the individual phenomena are discussed in more detail.

\subsection{Flares}\label{subsect:act_flares}

Stellar flares are the result of magnetic reconnection in the upper atmosphere which 
leads to particle acceleration and plasma heating. 
Footpoints of magnetic loops light up in optical/UV emission lines
and continuum following electron bombardment of the lower atmosphere.
Intense X-ray emission goes along with such events, and
is ascribed to the filling of the coronal loops with heated plasma evaporated off the
chromospheric layers along the magnetic field lines 
\citep[`chromospheric evaporation';][]{Antonucci84.1}.

{\em XMM-Newton} studies of flares comprise both detailed analyses of individual events
on nearby stars and statistical investigations of larger samples. An interesting capability 
of {\em XMM-Newton} for flare science is the simultaneous availability of the Optical Monitor 
which enables recording the white-light flare jointly with the X-ray signal. 

Some of the notable case studies carried out with {\em XMM-Newton}
regard huge flares on our nearest stellar neighbor, 
Proxima Cen \citep{Guedel04.1} and on the nearby M star CN\,Leo \citep{Liefke10.0}. 
The high photon statistics of these events allowed to establish a density
increase during the flare through a time-resolved analysis of the high-resolution RGS spectrum. 
The changing profile of the O\,{\sc VII} triplet during these flares 
indicates a temporary increase in the electron
density during phases which correspond to peaks in the X-ray luminosity.

For some X-ray flares, including the above-mentioned events, the optical counterpart has
been caught thanks to the simultaneous observation carried out with the Optical Monitor 
onboard {\em XMM-Newton}. A time-lag of few tens of seconds was determined by 
\cite{Stelzer06.4} 
between the peaks of the $V$ band and the EPIC/pn lightcurves for a giant flare on a 
late-M star. Such observations have substantially increased our knowledge of stellar
flare physics, and have confirmed some essential points in the canonical picture of a 
solar flare \citep{Cargill83.1}, providing support for the {\em solar-stellar analogy}. 

One point where the analogy of solar and stellar flares hits shaking ground regards 
the size of the X-ray emitting loop structures. The length of flaring plasma loops
can be inferred by adapting hydrodynamical (HD) models \citep{Reale97.1} to the
observed time-evolution of X-ray temperature and emission measure. 
Contrary to the case of the Sun where magnetic
loops cover only a small fraction of the surface \citep{Reale10.0}, coronal loops
on the order of a stellar radius have been derived for M dwarf stars 
\citep[e.g.,][]{Stelzer06.4}. Contrasting results have been presented from 
{\em Chandra} and {\em XMM-Newton} observations of flares on 
pre-MS stars: next to loops
of regular size \citep[e.g.][]{Franciosini07.0}, very large structures were derived for 
some flares and it was suspected that they represent magnetic loops 
that connect from the star to the circumstellar disk 
\citep{Favata05.1}. However, after the evolutionary stage of the individual objects 
was better constrained with the help of {\em Spitzer} data, 
the evidence for X-ray emitting star-disk loop structures has weakened 
\citep{Getman08.2, Aarnio10.0}. As an independent method to determine flare loop sizes,
\cite{LopezSantiago16.0} applied a wavelet-based analysis to brightness oscillations 
seen during some X-ray flares of TTS. 
Such oscillations had first been discovered in {\em XMM-Newton} observations of the 
field M dwarf AT\,Mic \citep{MitraKraev05.1}. Under the assumption that the oscillation
is caused by a wave within a magnetic flux tube, the observed period is through the
sound speed directly connected with the loop length. This analysis confirmed the 
large values obtained previously with HD models from the same data. A drawback of this
method is that it is applicable only to the small subset of flares which display 
detectable oscillations.  

Given the low occurrence rate of large flares 
\citep[$\sim 1$/yr for events of $E_{\rm F} > 10^{34}$\,erg for the Sun;][]{Schrijver12.0}, 
statistical studies of stellar X-ray flares require access to a (possibly) homogeneous database
comprising large numbers of stars. 
Such data, potentially, allow to characterize the number distributions of flares 
which are a crucial diagnostic for the importance of nano-flares in the coronal heating process. 
Flare energy number distributions have been extensively studied for the Sun
across a wide wavelength range from the UV to the hard X-ray band 
\citep[see e.g. references in][]{Aschwanden14.0}, but are much harder to come by for other stars.  
Lately, the {\em XMM-Newton Serendipitous Source
Catalog} \citep{Rosen16.0} 
has revealed to represent a valuable source for such studies. Using this database, 
\cite{Pye15.0} have identified $\sim 130$ X-ray flares on stars comprising spectral 
types F to M. A critical point in such studies is to take proper account of the survey
incompleteness which is difficult to assess in a non-uniform database. Here, studies of flares
from pre-MS stars offer an advantage, as X-ray observations
in star forming regions naturally provide data for many stars in one shot and all stars
are observed at the same flux limit. Respective systematic investigations in various
star forming regions suggest that the X-ray flare energies of TTS
follow a power-law with spectral index $\approx 2$ 
\citep{Stelzer07.1, Albacete07.1, Caramazza07.1}, similar to the case of the Sun.
However, applying this result to the regime of nano-flares requires an extrapolation
over many orders of magnitude in flare energy, since the pre-MS X-ray flare energy 
distributions are complete only for energies $\gtrsim 10^{35}$\,erg/s.  

The order of magnitude increase of the X-ray flux during stellar flares entails
that such events
can also be seen on stars with quiescent flux below the detection limits of current X-ray
instrumentation. In particular this applies to the ultracool dwarfs (UCDs), objects at the faint
end of the stellar sequence defined as having spectral type M7 and later which may be 
stars or brown dwarfs depending on the -- generally unknown -- age. The activity of UCDs
is gaining interest as some objects of this class have been shown to present circularly
polarized radio bursts similar to those of solar-system planets 
\citep{Burgasser05.1, Hallinan06.1}, 
indicative of a transition between stellar-like and planetary-like coronal properties in the
UCD regime. Based on the small sample of objects with multi-wavelength measurements, 
X-ray flaring and radio bursting seems to be mutually exclusive 
suggesting the presence of two classes of UCDs harboring distinct coronal conditions 
\citep{Stelzer12.0}. The upcoming {\em Evolutionary Map of the Universe (EMU)} to be
produced by the SKA precursor ASKAP \citep{Norris11.0} will likely represent a step-change 
in the number of radio detections of UCDs, 
offering many targets suitable for {\em XMM-Newton} follow-up with the aim to 
examine the above-mentioned dichotomy. 
\begin{figure}
\hspace*{1cm}
\includegraphics[width=6.0cm]{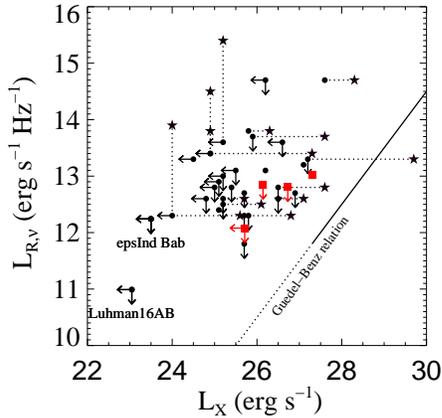}
\caption{X-ray versus radio luminosity for UCDs showing the violation of the G\"udel-Benz
relation for all other types of active stars. 
Flares are marked with star symbols. The populations of X-ray and radio flaring 
UCDs are largely disjoint. Stelzer et al., in prep.}
\label{fig:ucd_radio_xrays}
\end{figure}

\subsection{Rotational modulation}\label{subsect:act_rotmod}

Star spots are responsible for the most frequent and well-studied stellar variability 
phenomenon at optical wavelengths. 
The periodic brightness variation produced by surface spots moving in and
out of the line-of-sight in the course of a stellar rotation is also an excellent 
means to measure stellar rotation rates \citep[see e.g.][]{Herbst00.2}. 
In the X-ray band, the rotational modulation associated with structured coronae  -- if present
at all -- is usually masked by much stronger irregular variability related to flaring. 
As a consequence only isolated reports on periodic X-ray variability from
late-type stars are found in the literature: 
The most notable examples, both identified with {\em XMM-Newton}, are
VXR\,45 in the IC\,2391 star cluster \citep{Marino03.1} 
and the benchmark young field dwarf AB\,Dor \citep{Hussain07.1}. 
Interestingly, both stars have
similar spectral type (late-G; early-K), age ($\sim 50$\,Myr) and rotation rate 
($P_{\rm rot} = 0.2; 0.5$\,d). Fast rotation is known to be associated with larger
{\it optical} spot amplitude \citep[e.g.][]{McQuillan14.0, Stelzer16.0}. 
In this vein, the detection of X-ray
rotational modulation exclusively on stars with short rotation period may not be surprising. 

The more typical, longer rotational time-scales of late-type stars ($\sim 1-20$\,d) 
make it generally 
difficult to follow a whole star spot period in the X-ray band. A few star forming
regions have been observed with {\em XMM-Newton} for exposure times approaching the 
stellar rotational timescales \citep[e.g. $\rho$\,Oph;][]{Sciortino06.1} 
but no X-ray rotational modulation was reported so far. However, a systematic 
exploitation of these data has not yet taken place. Here, the analysis tools provided
by the EXTraS project\footnote{{\it Exploring the X-ray Transient and variable Sky} (EXTraS), 
funded within the EU seventh Framework Programme, is aimed at the thorough characterization 
of the variability of X-ray sources in archival {\em XMM-Newton} data; \cite{deLuca14.0}.} 
may prove useful. X-ray periodogram 
analysis may follow the example of \cite{Flaccomio05.1} who found X-ray periods 
in $\sim 10$\,\% of the stellar sample in a nearly uninterrupted $13\,{\rm d}$-long 
{\em Chandra} observation of the Orion Nebula Cluster (project `COUP'). 
Statistical methods examining changes of the amplitude of X-ray lightcurves throughout the 
phases of known periodic optical variability patterns can also be used to 
establish the presence of rotational modulation in the X-ray band \citep{Flaccomio12.0}.

\subsection{Dynamo cycles}\label{subsect:act_dyncyc}

Dynamo cycles, similar to the Sun's $11$-yr cycle, have been observed on $\sim 60$\,\% of 
solar-type MS stars through variability of the Ca\,{\sc ii}\,H\&K index 
\citep{Baliunas95.2}.
For a much smaller number of late-type stars, dynamo cycles have been identified 
through photometric monitoring \citep[e.g.][]{Olah09.0}, and only four stars
have established X-ray cycles. Three of the stars with X-ray cycles are binaries
\citep[see][and references therein]{Robrade12.0, Favata08.0, Ayres14.0}. 
 All these stars have long rotation periods, estimated ages of a few Gyrs 
 \citep{Barnes07.1}, 
 and intermediate to low Ca\,{\sc ii}\,H\&K activity levels. This is in line with results from 
 chromospheric studies, where activity cycles were found mostly among intermediate-active 
 stars while very active stars (in terms of $R_{\rm HK}$) tend to show irregular 
 Ca\,{\sc ii}\,H\&K variability \citep{Baliunas95.2}. 

\begin{figure}
\includegraphics[width=5.5cm, angle=90]{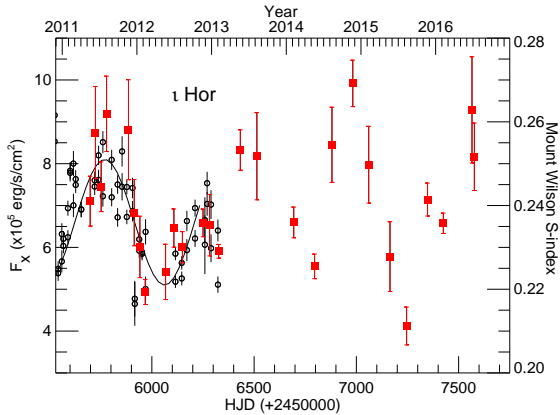}
\caption{X-ray and Ca\,{\sc ii} lightcurve of $\iota$\,Hor. Black open 
symbols -- Ca\,{\sc II} S-index, recent unpublished Ca\,{\sc II} data
is not shown; red -- {\em XMM-Newton} X-ray flux.  
Sanz-Forcada et al., in prep.}
\label{fig:iotahor}
\end{figure}

 \cite{SanzForcada13.0} reported 
the first detection of an X-ray activity cycle in a young solar analog: $\iota$\,Hor has
a spectral type of G0\,V and an age of $\sim 600$\,Myr \citep{Barnes07.1}. 
Its monitoring with {\em XMM-Newton} started in 2011, stimulated by the detection 
of a cycle in Ca\,{\sc ii}\,H\&K emission with only $1.6$\,yr 
period \citep{Metcalfe10.0}. The X-ray luminosity was soon seen to mimic the 
cyclic variation of the Ca\,{\sc ii} S-index but an interruption of the cycle in both
diagnostics seems to have taken place in 2013 with a subsequent phase shift 
when the periodic behavior resumed (see Fig.~\ref{fig:iotahor}).
$\iota$\,Hor can be considered a proxy for the Sun at an early evolutionary phase. 
Its age corresponds to the time when life developed on Earth, 
a process which might have been affected by the Sun's high-energy radiation 
\citep[see e.g.][]{Cnossen07.0}. 
Clearly, differences between X-ray cycles of young and old stars are 
relevant for our understanding of the early evolution of the Earth.
Whether the violent X-ray and Ca\,{\sc ii} cycle of $\iota$\,Hor is typical 
for dynamo activity of young Suns, therefore, is to be demonstrated through 
studies of other, similarly young stars. 

Evidently, decade-long monitoring is not easily feasible
in the X-ray band. The four known X-ray cycles have been constructed on the basis of
repeated snapshot observations over the years, sometimes combining data from different
instruments. As described e.g. by \cite{Ayres09.1} this requires careful cross-calibration,
and a long-term timeseries of a single instrument is certainly preferable. The lifetime
of {\em XMM-Newton} now extends over the typical duration of stellar activity cycles
and is perfectly suited for a systematic search of the associated X-ray variability.

Next to the stars for which dedicated long-term monitoring campaigns are already ongoing, 
indirect evidence for cycles may be collected from sparser multi-epoch observations. 
Rosen et al.\footnote{http://www.cosmos.esa.int/documents/332006/1018949/SRosen\_p.pdf} 
have reported on this conference that multiple visits of {\em XMM-Newton}
are available for $> 70000$ sources, with individual cases having 
up to $\sim 50$ observations. A preliminary sample based on positional matching with the 
CDS SIMBAD database yields a total of $\sim 600$ cool stars among the 
{\em XMM-Newton} sources having at least three separate observations;   
$\sim 10$\,\% of these show flux variations of a factor of $\gtrsim 5$ between different
observations and in many cases there is no discernible 
contribution from short-term variability 
(Pye et al. 2016\footnote{Proceedings 19th Cambridge Workshop on Cool Stars, Stellar 
Systems and the Sun, https://doi.org/10.5281/zenodo.58834}; J.P. Pye priv. comm.).

\section{X-ray variability from YSO environments}

Observations with {\em XMM-Newton} and {\em Chandra} led, 
in the early days of their operation, 
to the discovery that {\em Young Stellar Objects (YSOs)} 
can produce X-rays not only through their vigorous
coronae but also in shocks associated with disk accretion and protostellar jets. 
Since in- and outflows are dynamic processes, variability of the associated 
X-ray emission can be expected. 
Next to intrinsic variations, geometric effects may play a role in the 
observable (X-ray) emission as accretion streams 
or structures in the protostellar disk rotate in and out of the line-of-sight.   

X-ray signatures associated with the YSO environment are 
superposed on the -- often much stronger -- features produced by 
magnetic activity. 
Thanks to the great leap in spectral and spatial resolution provided by the instruments
onboard {\em XMM-Newton} and {\em Chandra} 
it became possible to disentangle these different X-ray emission mechanisms.

\subsection{Accretion}\label{subsect:yso_acc}

High-resolution X-ray spectra with the gratings onboard both satellites allowed for the
first time to examine individual emission lines in the soft X-ray band on a star other than
the Sun. This gave access to the flux ratios of helium-like triplets which 
represent a diagnostic for the density 
of the X-ray emitting plasma in cool stars \citep[see e.g.][]{Ness01.1}.
{\em XMM-Newton} emission line measurements provided evidence for high 
densities ($n_{\rm e} \sim 10^{13}\,{\rm cm^{-3}}$) 
in a handful of classical (i.e. accreting) 
TTS (see Sect.~\ref{sect:intro}) 
and indications for an extra soft X-ray component  
in these stars \citep{Guedel07.4}. Both phenomena represent the typical conditions in 
accretion shocks on low-mass stars. 

Few studies of X-ray accretion diagnostics have been carried out in the time-domain.
Only for the prototypical classical TTS TW\,Hya some evidence for temporarily 
enhanced mass accretion has been identified in soft X-ray emission lines observed with
{\em Chandra} and likely produced in the accretion shock \citep{Dupree12.0}. 
The observed variability supports the picture of an 
``accretion-fed corona" where the stellar X-ray emission is enhanced as a result of 
heating by the accreting matter \citep{Brickhouse12.0}. 

The changing visibility of magnetically channeled, accreting matter streams throughout a 
stellar rotation expected from magneto-hydrodynamic simulations 
(Romanova et al. 2008; Kurosawa et al. 2008) suggests that the observable X-ray accretion
signatures of pre-MS stars vary across a rotation cycle. 
However, due to the considerable observing time investment required to follow accreting stars 
during a full rotation, 
little is known on this end from the observational prospective. 
A notable exception is V4046\,Sgr, a binary composed of two young, accreting solar-mass stars 
in a synchronized $2.4$\,d orbit. With a $360$\,ks-long {\em XMM-Newton}/RGS pointing 
\cite{Argiroffi12.0} discovered the flux of the cool 
(accretion-driven) X-ray emission lines to vary periodically across the orbital/rotational 
cycle. The maxima were aligned with orbital phases in which the binary 
was seen in quadrature, indicating that the X-ray emitting accretion streams are 
located in the interbinary space. 
A clear periodic X-ray signal on a single accreting star was observed so far only 
on the Class\,I protostar, V1647\,Ori, an EX\,Or type variable.  
From multiple epochs of X-ray data involving {\em Chandra}, {\em XMM-Newton} and 
{\em Suzaku} observations, 
\cite{Hamaguchi12.0} modelled the phase-folded X-ray lightcurve with 
emission from accretion hot spots. 

EXOr and FUOr phenomena are sporadic and poorly understood brightness bursts 
(up to $4-5$\,mag in optical light), considered to 
represent the most extreme accretion events during the YSO phase 
and responsible for assembling a significant amount of the final stellar mass 
\citep{Hartmann96.1}. 
These events are classified according to the timescales of the
outbursts as FU\,Ors \citep[duration of decades;][]{Hartmann85.1} 
or as EX\,Ors \citep[duration of months and recurrent;][]{Herbig89.1}. 
Both types of outbursts are ascribed to drastic accretion rate increases 
($\dot{M}_{\rm acc} \sim 10^{-5...-6}\,{\rm M_\odot/yr}$), 
but the physical mechanisms triggering the outburst are still debated;  
see \cite{Audard14.0} for a recent review.
While the number of known EXOr and FUOr events is small by itself, even fewer
have been observed in X-rays and less than a handful have multiple epoch X-ray data. 
Several snapshots over the course of recent outbursts on V1118\,Ori and EX\,Lup 
involving both {\em XMM-Newton} and {\em Chandra} CCD cameras 
showed that the soft X-ray spectral component evolved towards lower flux, as expected
while accretion was fading, but the X-ray temperatures are too high for emission
from accretion shocks \citep[e.g.][]{Audard10.1, Teets12.0}. 

The number of known young eruptive stars is rapidly increasing as manifested by the 
large number of ATel notifications produced by programs such as 
EXORCISM \citep{Lorenzetti09.0}. This provides high potential 
for X-ray identifications of newly discovered events
and for follow-up during their evolution with repeated {\em XMM-Newton} pointings on
months to years time-scales, shedding light on the X-ray production mechanism during
violent accretion events on YSOs.

\begin{figure}[t]
\includegraphics[width=7.5cm]{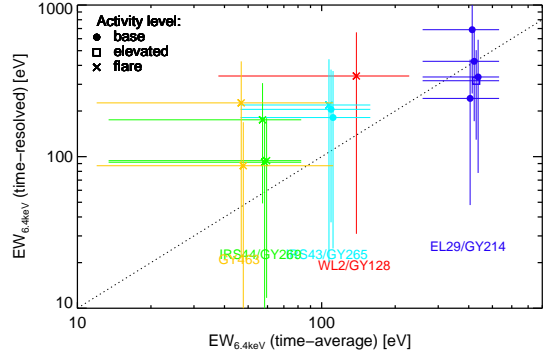}
\caption{Equivalent width (EW) of $6.4$\,keV emission in individual time segments of 
time-resolved spectroscopy vs. EW in the time-averaged spectrum of
YSOs from the DROXO project; \cite{Stelzer11.0}.}
\label{fig:droxo_feka}
\end{figure}

\subsection{$6.4$\,keV line emission} 

Among the most prominent feedback phenomena between stellar high-energy emission and 
the circumstellar matter is the $6.4$\,keV Fe\,K$\alpha$ line. 
This line, observed in some dozens of YSOs so far, can not be attributed to 
highly ionized coronal material \citep[e.g.][]{Tsujimoto05.1}. 
In the Sun the $6.4$\,keV emission represents K$\alpha$ fluorescence of neutral iron in 
the photosphere irradiated with X-rays from the solar corona during flares 
\citep{Bai79.0}. 
In YSOs this line is usually attributed to fluorescence from neutral iron 
{\it in the disk} following X-ray illumination from the star's corona. 
In some cases the line equivalent widths were shown to be incompatible with 
fluorescence calculations \citep{Giardino07.2, Czesla07.2}.  
Electron impact ionization in star-disk magnetic tubes was suggested as
an alternative excitation mechanism for the line. 
Another possible scenario that might explain 
large observed equivalent widths without the need to invoke 
collisional excitation is the (partial) occultation of the illuminating continuum flux, 
e.g. a stellar flare behind the limb \citep{Drake08.1}. 

The identification of the 
origin of the Fe K$\alpha$ line in YSOs can greatly benefit from time-resolved spectroscopy 
that enables a direct evaluation of the continuum emission visible at the moment when the line 
is produced. Several tens of ks exposure time with {\em XMM-Newton} are typically required
to obtain enough signal in the high-energy tail of a YSO X-ray spectrum for a detection
of the Fe\,K$\alpha$ line which is partly blended with the coronal line complex 
produced by highly ionized iron.
Time-resolved studies of the $6.4$\,keV line have been carried out for the Orion Nebula
cluster based on the 
 {\em Chandra} `COUP' project \citep{Czesla10.0} and for $\rho$\,Oph based on 
 {\em XMM-Newton}'s `DROXO' project. 
For the case of DROXO, \cite{Stelzer11.0} showed that higher equivalent widths are measured
for the Fe\,K$\alpha$ line in time-resolved analysis as compared to the time-averaged
study (see Fig.~\ref{fig:droxo_feka}). 
One question to follow up in future X-ray studies regards the indications for $6.4$\,keV 
emission in Class\,III objects, i.e. TTS 
that are believed to be devoid of circumstellar material such that the origin of the line
remains unclear.

\acknowledgements
Fig.3 is shown with kind permission of Astronomical Society of the Pacific 
Conference Series (ASPCS).

\bibliographystyle{an}
\bibliography{stelzer}

\end{document}